\documentclass[pre,numbers,amsmath,fleqn,amssymb]{revtex4}

\usepackage{graphicx}
\usepackage{amssymb}
\usepackage{dcolumn}
\usepackage{bm}

\begin{document}

\title{Wigner-Moyal description of free variable mass Klein-Gordon fields} 
\author{J. P. Santos}
\author{L. O. Silva} 
\email{luis.silva@ist.utl.pt}
\affiliation{GoLP/Centro de F\'isica dos Plasmas,  Instituto Superior T\' ecnico, 1049-001 Lisboa, Portugal}

\begin{abstract}
A system of coupled kinetic transport equations for the Wigner distributions of a free variable mass Klein-Gordon field is derived. This set of equations is formally equivalent to the full wave equation for electromagnetic waves in nonlinear dispersive media, thus allowing for the description of broadband radiation-matter interactions and the associated instabilities. The standard results for the classical wave action are recovered in the short wavelength limit of the generalized Wigner-Moyal formalism for the wave equation.
\end{abstract}

\maketitle
\section{Introduction}
The Wigner-Moyal formalism provides an alternative formulation for non-relativistic Quantum Mechanics, where the wave function is replaced by a quasi-particle distribution function, the Wigner distribution, in phase-space, which evolves in time according to a transport equation for the Wigner-Moyal distribution \cite{bib:wigner}. One of the main advantages of the Wigner-Moyal formalism is the possibility to describe quantum fields with non-trivial statistical properties \cite{bib:besieris1}.

Since the evolution of electromagnetic waves can be described, in the paraxial approximation, by a Schr\"odinger like equation is not surprising that the Wigner-Moyal formalism has been used to describe waves propagating in an inhomogeneous, dispersive, anisotropic and slowly varying in time medium \cite{bib:besieris}.
However, the propagation of electromagnetic waves is, in general, a two mode problem as clearly stated by the second order time derivative in the wave equation for electromagnetic waves in any medium \cite{bib:jackson}. Previous works \cite{bib:prev} have only dealt with the single mode problem, formally equivalent to propagating a nonlinear Schr\"odinger-like field, but this approach clearly breaks down when backscattering of the field is an important ingredient for the full dynamics of the system, as it usually occurs in laser/radiation-matter interactions at high intensities.

In this paper, we build an alternative description for the propagation of electromagnetic waves in dispersive and nonlinear media, capable of capturing both forward and backscattering dynamics of the electromagnetic waves, by generalizing the work developed by J. Javanainen \emph{et. al} \cite{bib:serimaa} to a variable mass problem. In the short wavelength approximation the results regarding the classical wave action (e. g. \cite{bib:mcdonald}) are recovered.

We first construct a $2\times2$ Wigner matrix \cite{bib:serimaa} on the basis of the Hamiltonian form of the Klein-Gordon equation of a charged scalar particle field, introduced by Feshbach and Villars \cite{bib:feshbach}. The diagonal elements describe forward and backward photon densities and off-diagonal elements correspond to cross-densities in phase-space. In the corresponding quantum problem, the mass is assumed to be fixed \cite{bib:serimaa}; here a further generalization is performed in order to study the variable mass problem, since the response of the medium (in our case, a cold plasma) exhibits spatial and temporal dependencies. The motivation of this approach is to develop the relativistic phase-space description of the dynamics of Klein-Gordon particles analogously to the Hamiltonian description used earlier by Bialynichi-Birula, G\'ornicki and Rafelski for the Dirac particles \cite{bib:rafelski}. The advantage of this method, compared with the manifestly covariant descriptions of \cite{bib:vasak}, is that a single time parameter is used to describe the dynamics; as a result, the phase-space is a genuine dynamical system with its fully prescribed dynamics, and its conservation laws.

In order for the discussion to be self-contained, a brief outline of the Feshbach and Villars Hamiltonian formulation for the Klein-Gordon equation is given in Section \ref{sec:two}, along with the definition of the $2\times2$ Wigner matrix. In Section \ref{sec:three}, we present the derivation of the generalized Boltzmann-Vlasov equation of motion for the Wigner matrix, and the set of coupled equations for the four real phase-space densities, which are related to the diagonal and cross-densities of the Klein-Gordon particles and their antiparticles, are derived. A physical interpretation of the phase-space densities is provided. In Section \ref{sec:four} the short wavelength approximation of the Boltzmann-Vlasov equation of motion for the Wigner matrix is performed, and the well known results of the classical wave action (e. g. \cite{bib:mcdonald}, and references therein) are recovered. Finally, in Section \ref{sec:five}, we state our conclusions.

\section{\label{sec:two}The Feshback-Villars formalism for electromagnetic waves and the Wigner matrix}

The propagation of an electromagnetic wave in a dispersive medium is described by the Klein-Gordon equation \cite{bib:jackson}:
\begin{equation}
\varepsilon^2 \partial_t^2 \Phi-\varepsilon^2 c^2 \overrightarrow{\nabla}_{\mathbf{r}}^2 \Phi+\omega^2_p(\mathbf{r},t)\Phi=0
\label{eqn:1}
\end{equation}
Here, $\Phi$ is stands for one of the spatial components of the vector potential, in the Coulomb gauge, and $c$ is the speed of light in vacuum. 
The CGS system of units is assumed throughout the paper, except where explicitly noted. We observe that $\omega_p$ plays the role of the mass of the Klein-Gordon field. For a cold plasma $\omega_p(\mathbf{r},t)$, is the electron plasma frequency, a function of time and space, given by
 $\omega_p(\mathbf{r},t) = \left(4 \pi e^2 n_e/m_e \gamma\right)^{1/2}$
where $e$ is the electron charge,  $m_e$ is the electron mass,  $n_e(\Phi(\mathbf{r},t),\mathbf{r},t)$ is the electron density in the plasma, and $\gamma(\Phi(\mathbf{r},t),\mathbf{r},t)$ is the relativistic Lorentz factor of the plasma electrons. The electron plasma frequency can be an arbitrary function of both time and space, and it may be a source of nonlinearity \cite{bib:kruer}. Equivalently, Eq.(\ref{eqn:1}) is also the starting point for nonlinear optics \cite{bib:shen}, 
which traditionally uses the index of refraction instead of $\omega_p$ to describe the properties of the media.  

A distinguishing feature of this problem is the presence of a positive dimensionless parameter $\varepsilon$. This parameter can be taken inversely proportional to the scale size of the spatial inhomogeneities, being small, but finite, for  a slowing varying medium. We stress that our discussion is valid for $\varepsilon=1$, which corresponds to the usual wave equation. The inclusion of this small parameter will allow us to easily perform the analysis in the short-wavelength limit in Section \ref{sec:four}.

To find the Hamiltonian form of Eq.~(\ref{eqn:1}), we follow the Feshbach-Villars \cite {bib:feshbach} description, defining:
\begin{equation}
\phi,\chi=\frac{1}{\sqrt{2}}\left(\Phi\pm\frac{i\varepsilon}{\omega_{p0}}\partial_t \Phi\right)
\label{eqn:2}
\end{equation}
where $\omega_{p0}^2$ is the background electron plasma frequency, independent of time and space, such that $\omega_p^2(\mathbf{r},t)=\omega_{p0}^2+\tilde{\omega}_p^2(\mathbf{r},t)$. Eq.~(\ref{eqn:1}) can then be written as:
\begin{subequations}
\begin{equation}
i\varepsilon\partial_t \phi=-\frac{\varepsilon^2c^2}{2\omega_{p0}}\overrightarrow{\nabla}_{\mathbf{r}}^2(\phi+\chi)+\frac{\tilde{\omega}_p^2}{2\omega_{p0}}(\phi+\chi)+\omega_{p0}\phi
\end{equation}
\begin{equation}
i\varepsilon\partial_t \chi=\frac{\varepsilon^2c^2}{2\omega_{p0}}\overrightarrow{\nabla}_{\mathbf{r}}^2(\phi+\chi)+\frac{\tilde{\omega}_p^2}{2\omega_{p0}}(\phi+\chi)-\omega_{p0}\chi
\end{equation}
\label{eqn:3}
\end{subequations}
There are several ways of separating equation (\ref{eqn:1}) in two coupled equations, respecting definitions (\ref{eqn:2}). 
Equation (\ref{eqn:3}) denotes the correct expansion for the scalar potential, $\tilde{\omega}_p^2$, and guarantees, unlike the prescription proposed in \cite{bib:serimaa}, that the short wavelength limit leads to the standard classical wave action conservation, as shown in Section IV.

If we now introduce $\Psi$, the two-component vector potential, as
\begin{equation}
\Psi=\left[
\begin{array}{c}
\phi
\\
\chi
\end{array}
\right] \quad, 
\end{equation}
equation (\ref{eqn:3}) can be rewritten as:
\begin{equation}
i\varepsilon \partial_t \Psi=\frac{(\tau_3+i\tau_2)}{2\omega_{p0}} \left(-\varepsilon^2c^2\overrightarrow{\nabla}_{\mathbf{r}}^2+\tilde{\omega}_p^2\right)\Psi+\omega_{p0}\tau_3\Psi= \mathcal{H}_{\mathrm{m}}\Psi
\label{eqn:4}
\end{equation}
where $\mathcal{H}_{\mathrm{m}}$ is the Hamiltonian matrix. In equation (\ref{eqn:4}), and henceforth, the notation:
\begin{equation}
\begin{array}{ccc}
\tau_0=\left[\begin{array}{cc}
1&0\\
0&1
\end{array}
\right], & &
\tau_1=\left[\begin{array}{cc}
0&1\\
1&0
\end{array}
\right]
\\
\\
\tau_2=\left[\begin{array}{cc}
0&-i\\
i&0
\end{array}
\right], & &
\tau_3=\left[\begin{array}{cc}
1&0\\
0&-1
\end{array}
\right]
\end{array}
\end{equation}
is used to represent the Pauli matrices.

Equation (\ref{eqn:4}) is a Schr\"odinger-like equation. Thus, the usual Wigner formalism can be applied to a $2\times2$ Wigner matrix, instead of using the Wigner formalism on a scalar function. Defining the $2\times2$ Wigner matrix using the Feshbach-Villars representation \cite{bib:serimaa}:
\begin{equation}
W(\mathbf{k},\mathbf{r},t)\equiv\left(\frac{1}{2\pi\varepsilon}\right)^{3}\int_{\mathbb{R}^3}\mathrm{d}\mathbf{y}e^{i\frac{\mathbf{k}\cdot\mathbf{y}}{\varepsilon}}\Psi\left(\mathbf{r}_-,t\right)\overline{\Psi}\left(\mathbf{r}_+,t\right)
\label{eqn:5}
\end{equation}
where $\overline{\Psi}=\Psi^{\dagger}\tau_3$ denotes the Feshbach-Villars adjoint of the vector potential $\Psi$, $\mathbf{r}_{\pm}=\mathbf{r}\pm\mathbf{y}/2$, and $\mathbb{R}^3$ the usual three dimensional Cartesian space . By construction, $W$ is an hermitian matrix in the Feshbach-Villars sense, namely $\overline{W}\equiv\tau_3 W^{\dagger}\tau_3=W$, and its explicit form can be expressed in terms of the field components $\phi$ and $\chi$, yielding:
\begin{equation}
W=\left[
\begin{array}{cc}
W_{\phi\phi} & -W_{\phi\chi}^{*}
\\
W_{\phi\chi} & -W_{\chi\chi}
\end{array}
\right]
\end{equation}
where:
\begin{subequations}
\begin{equation}
W_{\phi\phi}=\left(\frac{1}{2\pi\varepsilon}\right)^{3}\int_{\mathbb{R}^3}\mathrm{d}\mathbf{y}e^{i\frac{\mathbf{k}\cdot\mathbf{y}}{\varepsilon}}\phi^*\left(\mathbf{r}_+,t\right)\phi\left(\mathbf{r}_-,t\right)
\end{equation}
\begin{equation}
W_{\phi\chi}=\left(\frac{1}{2\pi\varepsilon}\right)^{3}\int_{\mathbb{R}^3}\mathrm{d}\mathbf{y}e^{i\frac{\mathbf{k}\cdot\mathbf{y}}{\varepsilon}}\phi^*\left(\mathbf{r}_+,t\right)\chi\left(\mathbf{r}_-,t\right)
\end{equation}
\begin{equation}
W_{\chi\chi}=\left(\frac{1}{2\pi\varepsilon}\right)^{3}\int_{\mathbb{R}^3}\mathrm{d}\mathbf{y}e^{i\frac{\mathbf{k}\cdot\mathbf{y}}{\varepsilon}}\chi^*\left(\mathbf{r}_+,t\right)\chi\left(\mathbf{r}_-,t\right)
\end{equation}
\end{subequations}

It is useful to calculate the equation governing the time evolution of $\overline{\Psi}$. Taking the conjugate transpose of equation Eq.~(\ref{eqn:4}), and multiplying on the right by $\tau_3$, we obtain:
\begin{equation}
i\varepsilon \partial_t \overline{\Psi}=-\overline{\Psi}\frac{(\tau_3+i\tau_2)}{2\omega_{p0}} \left(-\varepsilon^2c^2\overleftarrow{\nabla}_{\mathbf{r}}^2+\tilde{\omega}_p^2\right)-\overline{\Psi}\tau_3\omega_{p0} =-\overline{\Psi}\mathcal{H}_m
\label{eqn:6}
\end{equation}
where we used $\tau_i^{\dagger}=\tau_i,\;i\in\{0,1,2,3\}$ and $\{\tau_i,\tau_j\}=2\tau_0\delta_{ij},\;i,j\in\{1,2,3\}$. Equation (\ref{eqn:6}) demonstrates that the Hamiltonian matrix is not hermitian. As a consequence the transport equation for the  Wigner matrix will have to include both commutators and anticommutators.

\section{\label{sec:three}Generalized Boltzmann-Vlasov equation of motion}

The equation of motion for the Wigner matrix (\ref{eqn:5}), can be derived from the Schr\"odinger-type equations (\ref{eqn:4}) and (\ref{eqn:6}) for, respectively, $\Psi\left(\mathbf{r}_-,t\right)$ and $\overline{\Psi}\left(\mathbf{r}_+,t\right)$, following a standard procedure \cite{bib:liboff}.

From equations (\ref{eqn:4},\ref{eqn:5},\ref{eqn:6}) we obtain
\begin{equation}
\partial_t W(\mathbf{k},\mathbf{r},t)=\mathcal{R}_1+\mathcal{R}_2-\frac{i}{\varepsilon}\omega_{p0}\left[\tau_3,W\right]
\label{eqn:7}
\end{equation}
where 

\begin{subequations}
\begin{equation}
\partial_t W(\mathbf{k},\mathbf{r},t)  =\left(\frac{1}{2\pi\varepsilon}\right)^{3}\int_{\mathbb{R}^3}\mathrm{d}\mathbf{y}e^{i\frac{\mathbf{k}\cdot\mathbf{y}}{\varepsilon}}\left[\partial_t \Psi(\mathbf{r}_-,t)\overline{\Psi}(\mathbf{r}_+,t)+\Psi(\mathbf{r}_-,t)\partial_t\overline{\Psi}(\mathbf{r}_+,t)\right]
\end{equation}
\begin{eqnarray}
\mathcal{R}_1 = \frac{i\varepsilon c^2}{\omega_{p0}}\left(\frac{1}{2\pi\varepsilon}\right)^{3}\int_{\mathbb{R}^3}\mathrm{d}\mathbf{y}e^{i\frac{\mathbf{k}\cdot\mathbf{y}}{\varepsilon}}\left\{(\tau_3+i\tau_2)\left[\frac{\overrightarrow{\nabla}_{\mathbf{r}}^2}{2}\Psi(\mathbf{r}_-,t)\right]\overline{\Psi}(\mathbf{r}_+,t)-\right. \nonumber \\
 \left.\Psi(\mathbf{r}_-,t)\left[\frac{\overrightarrow{\nabla}_{\mathbf{r}}^2}{2}\overline{\Psi}(\mathbf{r}_+,t)\right](\tau_3+i\tau_2)\right\}
\end{eqnarray}
\begin{eqnarray}
\mathcal{R}_2=\frac{i}{2\varepsilon\omega_{p0}}\left(\frac{1}{2\pi\varepsilon}\right)^{3}\int_{\mathbb{R}^3}\mathrm{d}\mathbf{y}e^{i\frac{\mathbf{k}\cdot\mathbf{y}}{\varepsilon}}\left[\Psi(\mathbf{r}_-,t)\overline{\Psi}(\mathbf{r}_+,t)(\tau_3+i\tau_2)\tilde{\omega}_{p_+}^2\right.\nonumber \\
-\left.(\tau_3+i\tau_2)\tilde{\omega}_{p_-}^2\Psi(\mathbf{r}_-,t)\overline{\Psi}(\mathbf{r}_+,t)\right]
\label{eqn:def}
\end{eqnarray}
\end{subequations}
where $\tilde{\omega}_{p_{\pm}}^2=\tilde{\omega}_{p}^2(\mathbf{r}_{\pm},t)$. Note that $\left[A,B\right]$ represents the usual commutator between operator $A$ and $B$, and that we also used the following identities:
\begin{equation}
\overrightarrow{\nabla}_{\mathbf{r}_{\pm}}g(\mathbf{r}_{\pm},t)=\overrightarrow{\nabla}_{\mathbf{r}}g(\mathbf{r}_{\pm},t)=\pm2\overrightarrow{\nabla}_{\mathbf{y}}g(\mathbf{r}_{\pm},t)
\label{eqn:8}
\end{equation}
where $g(\mathbf{r},t)$ stands for a first order differentiable function of $\mathbf{r}$, and an arbitrary function of $t$.

The first term on the right hand side of Eq.~(\ref{eqn:7}), $\mathcal{R}_1$,is the kinetic contribution for the equation of motion, and it is always present, independently of the potential $\tilde{\omega}_p^2(\mathbf{r},t)$, whereas the second term, $\mathcal{R}_2$, accounts for the the potential contribution. After some lengthy calculations $\mathcal{R}_1$, can also be written as:
\begin{equation}
\mathcal{R}_1=-\frac{c^2\mathbf{k}\cdot\overrightarrow{\nabla}_{\mathbf{r}}}{2\omega_{p0}}\left\{\tau_3+i\tau_2,W\right\}-\frac{i}{\varepsilon}\left[\mathcal{H}_0(\mathbf{k})(\tau_3+i\tau_2),W\right]+\mathcal{Q}
\end{equation}
where
\begin{subequations}
\begin{equation}
\mathcal{H}_0(\mathbf{k})=\frac{c^2}{\omega_{p0}}\mathbf{k}^2
\end{equation}
\begin{eqnarray}
\mathcal{Q}=-2\frac{i\varepsilon c^2}{\omega_{p0}}\left(\frac{1}{2\pi\varepsilon}\right)^3 \int_{\mathbb{R}^3}\mathrm{d}\mathbf{y}e^{i\frac{\mathbf{k}\cdot\mathbf{y}}{\varepsilon}}\left[(\tau_3+i\tau_2)\overrightarrow{\nabla}_{\mathbf{y}}\Psi(\mathbf{r}_-,t)\cdot\overrightarrow{\nabla}_{\mathbf{y}}\overline{\Psi}(\mathbf{r}_+,t)-\right.\nonumber
\\
\left.\overrightarrow{\nabla}_{\mathbf{y}}\Psi(\mathbf{r}_-,t)\cdot\overrightarrow{\nabla}_{\mathbf{y}}\overline{\Psi}(\mathbf{r}_+,t)(\tau_3+i\tau_2)\right]
\label{eqn:9}
\end{eqnarray}
\end{subequations}
and $\{A,B\}$ represents the usual anticommutator. Consequently, the equation of motion already denotes the effects of the non hermiticity of the Hamiltonian $\mathcal{H}_m$, as previously noted.

The form of equation (\ref{eqn:9}) suggests that a second order derivative in space, and a commutator must be involved; in fact, $\mathcal{Q}$ can be re-written as:
\begin{equation}
\mathcal{Q}=\frac{i\varepsilon c^2}{\omega_{p0}}\left[(\tau_3+i\tau_2)\frac{\overrightarrow{\nabla}_{\mathbf{r}}^2}{8},W\right]+\frac{i}{\varepsilon}\left[\frac{1}{2}\mathcal{H}_0(\mathbf{\hat{k}})(\tau_3+i\tau_2),W\right]
\end{equation}
leading to:
\begin{equation}
\mathcal{R}_1=-\frac{c^2\mathbf{k}\cdot\overrightarrow{\nabla}_{\mathbf{r}}}{2\omega_{p0}}\left\{\tau_3+i\tau_2,W\right\}-\frac{i}{\varepsilon}\left[\frac{1}{2}\mathcal{H}_0(\mathbf{\hat{k}})(\tau_3+i\tau_2),W\right]
\label{eqn:10}
\end{equation}
with the generalized wave number $\mathbf{\hat{k}}^2=\mathbf{k}^2-\varepsilon^2\frac{\overrightarrow{\nabla}_{\mathbf{r}}^2}{4}$

To calculate $\mathcal{R}_2$, we observe first that the inverse Fourier transform of the Wigner matrix (\ref{eqn:5}) yields the vector potential matrix $\Psi(\mathbf{r}_+,t)\overline{\Psi}(\mathbf{r}_-,t)$, \emph{viz.}:
\begin{equation}
\Psi(\mathbf{r}_+,t)\overline{\Psi}(\mathbf{r}_-,t)=\int_{\mathbb{R}^3}\mathrm{d}\mathbf{k}^{\prime}e^{-i\frac{\mathbf{k}^{\prime}\cdot\mathbf{y}}{\varepsilon}}W(\mathbf{k}^{\prime},\mathbf{r},t)
\label{eqn:temp1}
\end{equation}
Combining Eq.~(\ref{eqn:temp1}) and Eq.~(\ref{eqn:def}) leads to
\begin{eqnarray}
\mathcal{R}_2=\frac{i}{2\varepsilon\omega_{p0}}\left(\frac{1}{2\pi\varepsilon}\right)^{3}\int_{\mathbb{R}^3}\mathrm{d}\mathbf{k}^{\prime}\int_{\mathbb{R}^3}\mathrm{d}\mathbf{y}e^{i\frac{(\mathbf{k}-\mathbf{k}^{\prime})\cdot\mathbf{y}}{\varepsilon}}\left[W(\mathbf{k}^{\prime},\mathbf{r},t)(\tau_3+i\tau_2)\tilde{\omega}_{p_+}^2\right. \nonumber \\
\left.-(\tau_3+i\tau_2)W(\mathbf{k}^{\prime},\mathbf{r},t)\tilde{\omega}_{p_-}^2\right]
\end{eqnarray}
Since $\tilde{\omega}_{p_\pm}^2=e^{\pm i\frac{\mathbf{y}}{2i}\cdot\overrightarrow{\nabla}_{\mathbf{r}}}\tilde{\omega}_p^2(\mathbf{r},t)$, and in the case of a sufficient regular function $g$ \cite{bib:liboff}:
\begin{equation}
\int_{-\infty}^{+\infty}e^{i y k^{\prime}}g(iy)F(k^{\prime})\mathrm{d}k^{\prime}=\int_{-\infty}^{+\infty}e^{i y k^{\prime}}g\left(-\frac{\partial }{\partial k^{\prime}}\right)F(k^{\prime})\mathrm{d}k^{\prime}
\label{eqn:12}
\end{equation}
Thus, $\mathcal{R}_2$ can be expressed as:
\begin{equation}
\mathcal{R}_2=\frac{i\tilde{\omega}_p^2(\mathbf{r},t)}{2\varepsilon\omega_{p0}} \cos{\left(\frac{\varepsilon}{2}\overleftarrow{\nabla}_\mathbf{r}\cdot\overrightarrow{\nabla}_{\mathbf{k}}\right)}\left[W,(\tau_3+i\tau_2)\right]+\frac{\tilde{\omega}_p^2(\mathbf{r},t)}{2\varepsilon\omega_{p0}} \sin{\left(\frac{\varepsilon}{2}\overleftarrow{\nabla}_\mathbf{r}\cdot\overrightarrow{\nabla}_{\mathbf{k}}\right)}\left\{W,(\tau_3+i\tau_2)\right\}
\label{eqn:13}
\end{equation}

Combining Eqns.~(\ref{eqn:10},\ref{eqn:13}) in Eq.~(\ref{eqn:7}),  the generalized Boltzmann-Vlasov equation of motion for the Wigner matrix is:
\begin{eqnarray}
&&\partial_tW(\mathbf{k},\mathbf{r},t)+(\hat{D}-\hat{S})\frac{1}{2}\left\{\tau_3+i\tau_2,W(\mathbf{k},\mathbf{r},t)\right\}
\nonumber
\\
&&+\frac{i}{\varepsilon}\left[\mathcal{H}_0(\mathbf{\hat{k}})+\hat{C}\right]\frac{1}{2}\left[(\tau_3+i\tau_2),W(\mathbf{k},\mathbf{r},t)\right]+\frac{i\omega_{p0}}{\varepsilon}\left[\tau_3,W(\mathbf{k},\mathbf{r},t)\right]=0
\label{eqn:14}
\end{eqnarray}
where:
\begin{equation}
\hat{D}=\frac{c^2}{\omega_{p0}}\mathbf{k}\cdot\overrightarrow{\nabla}_\mathbf{r}, \qquad \hat{S}=\frac{\tilde{\omega}_p^2(\mathbf{r},t)}{\varepsilon\omega_{p0}} \sin{\left(\frac{\varepsilon}{2}\overleftarrow{\nabla}_\mathbf{r}\cdot\overrightarrow{\nabla}_{\mathbf{k}}\right)}, \qquad \hat{C}=\frac{\tilde{\omega}_p^2(\mathbf{r},t)}{\omega_{p0}} \cos{\left(\frac{\varepsilon}{2}\overleftarrow{\nabla}_\mathbf{r}\cdot\overrightarrow{\nabla}_{\mathbf{k}}\right)}
\label{eqn:trig}
\end{equation}
In the argument of the trigonometric functions, the operator $\overleftarrow{\nabla}_{\mathbf{r}}$ acts on the electron plasma frequency, and the operator $\overrightarrow{\nabla}_{\mathbf{k}}$ acts on the components of the Wigner matrix $W(\mathbf{k},\mathbf{r},t)$. The matrix operator $\mathcal{H}_0(\mathbf{\hat{k}})$ reduces, in the short wavelength approximation, to the free particle Hamiltonian $\mathcal{H}_0(\mathbf{k})$. Both $\hat{C}$ and $\hat{S}$ can be expanded in a power series, containing only even powers of $\varepsilon$ and, therefore, $\hat{S}$ and $\hat{C}$ in equations (\ref{eqn:trig}) have regular limits when $\varepsilon\to0$.

The operators (\ref{eqn:trig}) can also be cast in an integral form, which is more useful for numerical implementations:
\begin{equation}
\hat{C}, \hat{S} f(\mathbf{k},\mathbf{r},t)=\int_{\mathbb{R}^3}\mathrm{d}\mathbf{k}^{\prime} \mathcal{K}^{\pm} (\mathbf{k}-\mathbf{k}^{\prime},\mathbf{r},t)f(\mathbf{k}^{\prime},\mathbf{r},t)
\end{equation}
where
\begin{equation}
\mathcal{K}^{\pm}(\mathbf{k},\mathbf{r},t)=\frac{i}{2\varepsilon\omega_{p0}}\left(\frac{1}{2\pi\varepsilon}\right)^{3}\int_{\mathbb{R}^3}\mathrm{d}\mathbf{y}e^{i\frac{\mathbf{k}\cdot\mathbf{y}}{\varepsilon}}\left(\tilde{\omega}_{p_+}^2-\tilde{\omega}_{p_-}^2\right)
\end{equation}

Equation (\ref{eqn:14}) describes the equation of motion for the Wigner matrix, and therefore it describes the transport of $W$ in phase-space. Even though it is formally equivalent to the wave equation (\ref{eqn:1}), it is more useful to represent our field in terms of real Wigner distributions. Since the Pauli matrices form a basis for the $2\times2$ complex matrix space, $W$ admits the expansion 
\begin{equation}
W=\frac{1}{2}(\tau_0W_0+i \tau_1W_1-i\tau_2 W_2+\tau_3W_3)
\label{eqn:15}
\end{equation}
where all the $W_i,\;i\in\{0,1,2,3\}$ are real distributions. Since $W$ is not Hermitian, an expansion in the form $W=\sum_{i=0}^3 W_i\tau_i$, where all the $W_i,\;i\in\{0,1,2,3\}$, are real functions, is not possibe. Nevertheless, $W$ is Hermitian in the Feshbach-Villars sense and the natural expansion with real coefficients is given by Eq.~(\ref{eqn:15}). It is easily seen that $W_0=W_{\phi\phi}-W_{\chi\chi}$, $W_1=2 \mathrm{Im}(W_{\phi\chi})$, $W_2=2\mathrm{Re}(W_{\phi\chi})$ and $W_3=W_{\phi\phi}+W_{\chi\chi}$. The physical meaning of  $W_2(\mathbf{k},\mathbf{r},t)+W_3(\mathbf{k},\mathbf{r},t)$ is of interest, because it is related to the Wigner function of the vector potential $\Phi$. In fact, with 
\begin{equation}
W_{\Phi\Phi}(\mathbf{k},\mathbf{r},t)=\left(\frac{1}{2\pi\varepsilon}\right)^3  
 \int_{\mathbb{R}^3}\mathrm{d}\mathbf{y}e^{i\frac{\mathbf{k}\cdot\mathbf{y}}{\varepsilon}}\Phi^*\left(\mathbf{r}_{+},t\right)
\Phi\left(\mathbf{r}_{-},t\right)
\end{equation}
where $\Phi^*\left(\mathbf{r},t\right)$ stands for the complex conjugate of $\Phi\left(\mathbf{r},t\right)$, then 
\begin{equation}
W_{\Phi\Phi}=W_2(\mathbf{k},\mathbf{r},t)+W_3(\mathbf{k},\mathbf{r},t)= W_{\phi\phi}+W_{\chi\chi}+2\mathrm{Re}(W_{\phi\chi})
\end{equation}
The vector potential is completely characterised by the distribution function $W_{\Phi\Phi}$, which can be calculated for different electromagnetic field configurations. Furthermore, there is a one-to-one correspondence between the distribution $W_{\Phi\Phi}$, and the vector potential $\Phi$, apart from a constant phase shift $\phi_0$. Defining the inverse Fourier transform of $W_{\Phi\Phi}$, as:
\begin{equation}
\mathcal{F}^{-1}_{\mathbf{k},\mathbf{y}}(\mathbf{y},\mathbf{r},t)=\int_{\mathbb{R}^3}\mathrm{d}\mathbf{k}e^{-i\frac{\mathbf{k}\cdot\mathbf{y}}{\varepsilon}}\left[W_2(\mathbf{k},\mathbf{r},t)+W_3(\mathbf{k},\mathbf{r},t)\right]
\end{equation}
$\Phi(\mathbf{r},t)$ is determined by:
\begin{equation}
\Phi(\mathbf{r},t)=\frac{\mathcal{F}^{-1}_{\mathbf{k},\mathbf{r}}\left(\mathbf{r},\frac{\mathbf{r}}{2},t\right)}{\sqrt{\mathcal{F}^{-1}_{\mathbf{k},\mathbf{r}}(0,0,t)}}e^{i\phi_0}
\label{eqn:temp2}
\end{equation}
In Eq.~(\ref{eqn:temp2}) $\phi_0$ is determined from the initial conditions, and the inverse Fourier transform of the Wigner function is assumed to be different from $0$ at $t=0$.

The imaginary part of the cross-density $W_{\phi \chi}$, $W_1$, is the classical equivalent of the quantum 'Zitterbewegung' \cite{bib:serimaa}, and it represents the interference between forward and backward photons. For Wigner matrices of a free forward plane wave or a free backward plane wave, of momentum $\mathbf{k}_0$, $W_1$ is zero. 

As a simple illustration, let us consider a vector potential described by the superposition of two plane waves $\Phi(\mathbf{r},t)=\Phi_0\exp{\left[i(\mathbf{k}_0\cdot\mathbf{r}-\omega_0t)\right]}+\Phi_1\exp{\left[i(\mathbf{k}_1\cdot\mathbf{r}-\omega_1t)\right]}$, propagating in an uniform medium with $\tilde{\omega}_p^2(\mathbf{r},t)=0$. For this field:
\begin{subequations}
\begin{equation}
\phi(\mathbf{r},t)=\frac{\Phi_0}{2}\left(\frac{\omega_{p0}+\omega_0}{\omega_{p0}}\right)e^{i(\mathbf{k}_0\cdot\mathbf{r}-\omega_0t)}+\frac{\Phi_1}{2}\left(\frac{\omega_{p0}+\omega_1}{\omega_{p0}}\right)e^{i(\mathbf{k}_1\cdot\mathbf{r}-\omega_1t)}
\end{equation}
\begin{equation}
\chi(\mathbf{r},t)=\frac{\Phi_0}{2}\left(\frac{\omega_{p0}-\omega_0}{\omega_{p0}}\right)e^{i(\mathbf{k}_0\cdot\mathbf{r}-\omega_0t)}+\frac{\Phi_1}{2}\left(\frac{\omega_{p0}-\omega_1}{\omega_{p0}}\right)e^{i(\mathbf{k}_1\cdot\mathbf{r}-\omega_1t)}
\end{equation}
\end{subequations}

The corresponding components of the Wigner matrix are:
\begin{subequations}
\begin{eqnarray}
W_{\phi\phi}(\mathbf{k},\mathbf{r},t)  = \frac{(\omega_0+\omega_{p0})^2}{4\omega_{p0}^2}\Phi_0^2\delta(\mathbf{k}-\mathbf{k}_0)+\frac{(\omega_1+\omega_{p0})^2}{4\omega_{p0}^2}\Phi_1^2\delta(\mathbf{k}-\mathbf{k}_1)\nonumber
\\
+\frac{(\omega_{p0}+\omega_0)(\omega_{p0}+\omega_1)}{2\omega_{p0}^2}\Phi_0\Phi_1\cos{(\eta)}\delta\left(\mathbf{k}-\frac{\mathbf{k}_0+\mathbf{k}_1}{2}\right)
\end{eqnarray}
\begin{eqnarray}
W_{\phi\chi}(\mathbf{k},\mathbf{r},t) = -\frac{\mathbf{k}_0^2c^2}{4\omega_{p0}^2}\Phi_0^2\delta(\mathbf{k}-\mathbf{k}_0)-\frac{\mathbf{k}_1^2c^2}{4\omega_{p0}^2}\Phi_1^2\delta(\mathbf{k}-\mathbf{k}_1)+\frac{\Phi_0\Phi_1}{2}\left\{\left[1-\frac{\omega_0\omega_1}{\omega_{p0}^2}\right]\cos{(\eta)}\right.\nonumber \\
\left.+i\frac{\omega_0-\omega_1}{\omega_{p0}}\sin{(\eta)}\right\}\delta\left(\mathbf{k}-\frac{\mathbf{k}_0+\mathbf{k}_1}{2}\right)
\end{eqnarray}
\begin{eqnarray}
W_{\chi\chi}(\mathbf{k},\mathbf{r},t) = \frac{(\omega_0-\omega_{p0})^2}{4\omega_{p0}^2}\Phi_0^2\delta(\mathbf{k}-\mathbf{k}_0)+\frac{(\omega_1-\omega_{p0})^2}{4\omega_{p0}^2}\Phi_1^2\delta(\mathbf{k}-\mathbf{k}_1)\nonumber
\\
+\frac{(\omega_{p0}-\omega_0)(\omega_{p0}-\omega_1)}{2\omega_{p0}^2}\Phi_0\Phi_1\cos{(\eta)}\delta\left(\mathbf{k}-\frac{\mathbf{k}_0+\mathbf{k}_1}{2}\right)
\end{eqnarray}
\end{subequations}
where $\eta=\Delta_\mathbf{k}\cdot\mathbf{r}-\Delta_\omega t$, $\Delta_\mathbf{k}=\mathbf{k}_1-\mathbf{k}_0$, $\Delta_\omega=\omega_1-\omega_0$, $\delta(\mathbf{k})$ represents the Dirac delta distribution and $\omega_{0,1}=\sqrt{\mathbf{k}_{0,1}^2c^2+\omega_{p0}^2}$. The presence of two photon beams associated with the two plane waves is easily seen. Furthermore there is a "third wave" which results from the interference between the two plane waves. In this case, $W_1$ is different from zero, since it accounts for the interference between forward and backward photons. $W_0$ is associated with the effective photon density, \emph{i.e.}, it is positive when the forward photon density is bigger than the backward photon density.

Using the decomposition $W=\mathrm{Re}(W)+i\mathrm{Im}(W)$ in equation (\ref{eqn:14}), one can derive a set of two coupled real equations:
\begin{subequations}
\begin{eqnarray}
\partial_t 2\mathrm{Re}(W)+(\hat{D}-\hat{S})\frac{1}{2}\left\{\tau_3+i\tau_2,2\mathrm{Re}(W)\right\}-\frac{1}{\varepsilon}\left[\mathcal{H}_0(\mathbf{\hat{k}})+\hat{C}\right]\frac{1}{2}\left[(\tau_3+i\tau_2),\tau_1\right]W_1\nonumber
\\
-\frac{\omega_{p0}}{\varepsilon}\left[\tau_3,\tau_1\right]W_1=0
\end{eqnarray}
\begin{equation}
\partial_t (\tau_1W_1)+\frac{1}{\varepsilon}\left[\mathcal{H}_0(\mathbf{\hat{k}})+\hat{C}\right]\frac{1}{2}\left[(\tau_3+i\tau_2),2\mathrm{Re}(W)\right]+\frac{\omega_{p0}}{\varepsilon}\left[\tau_3,2\mathrm{Re}(W)\right]=0
\end{equation}
\label{eqn:16}
\end{subequations}
where we have used $\{\tau_i,\tau_j\}=2\tau_0\delta_{ij},\;i,j\in\{1,2,3\}$ and $2\mathrm{Im}(W)=\tau_1W_1$. From expansion (\ref{eqn:15}) it is easily seen that $2\mathrm{Re}(W)=\tau_0W_0-i\tau_2W_2+\tau_3W_3$ and, therefore, the four real phase-space densities obey the set of transport equations:
\begin{subequations}
\begin{equation}
\partial_t W_0+(\hat{D}-\hat{S})(W_2+W_3)=0
\end{equation}
\begin{equation}
\partial_t W_1-\frac{1}{\varepsilon}\left[\mathcal{H}_0(\mathbf{\hat{k}})+\hat{C}\right](W_2+W_3)-\frac{2}{\varepsilon}\omega_{p0}W_2=0
\end{equation}
\begin{equation}
\partial_t W_2-(\hat{D}-\hat{S})W_0+\frac{1}{\varepsilon}\left[\mathcal{H}_0(\mathbf{\hat{k}})+\hat{C}\right]W_1+\frac{2}{\varepsilon}\omega_{p0}W_1=0
\end{equation}
\begin{equation}
\partial_t W_3+(\hat{D}-\hat{S})W_0-\frac{1}{\varepsilon}\left[\mathcal{H}_0(\mathbf{\hat{k}})+\hat{C}\right]W_1=0
\end{equation}
\label{eqn:17}
\end{subequations}

The set of equations (\ref{eqn:17}) is the main result of this paper. With the appropriate definitions for $W_i$, the set of equations (\ref{eqn:17})  is formally equivalent to equation (\ref{eqn:1}), and therefore opens the way to an appropriate description of the propagation of electromagnetic waves with arbitrary spectral content in nonlinear dispersive media \cite{bib:santos}.

\section{\label{sec:four}The short wavelength limit}

We now examine the short wavelength limit of the Boltzmann-Vlasov equation of motion (\ref{eqn:14}) by examining the limit $\varepsilon\to0$. In order to consider the short wavelength limit of (\ref{eqn:14}) we first observe that:
\begin{equation}
\lim_{\varepsilon\to0}\hat{S}=\frac{\overrightarrow{\nabla}_{\mathbf{r}}\tilde{\omega}_p^2(\mathbf{r},t)}{2\omega_{p0}}\cdot\overrightarrow{\nabla}_{\mathbf{k}}, \qquad \lim_{\varepsilon\to0}\hat{C}=\frac{\tilde{\omega}_p^2(\mathbf{r},t)}{\omega_{p0}}, \qquad \lim_{\varepsilon\to0}\mathbf{\hat{k}}=\mathbf{k}
\label{eqn:18}
\end{equation}

Let the upperscript $^{(0)}$ represent any function in the short wavelength limit. In the limit $\varepsilon\to0$, the set of equations (\ref{eqn:16}) converges to a meaningful limit only if $W_1\to0$ faster than $\varepsilon\to0$; hence $W^{(0)}$ must be real, \emph{i.e.} $W_1^{(0)}=0$. The same conclusions can also be obtained using the coarse-graining technique introduced by Shin and Rafelski \cite{bib:shin}. Using the results in (\ref{eqn:18}) and the previous considerations, Eqs.~(\ref{eqn:16}) can be written, in the $\epsilon\to0$ limit, as:
\begin{subequations}
\begin{equation}
\partial_t W^{(0)}+\left[\frac{c^2}{\omega_{p0}}\mathbf{k}\cdot\overrightarrow{\nabla}_{\mathbf{r}}-\frac{1}{2}\frac{\overrightarrow{\nabla}_{\mathbf{r}}\tilde{\omega}_p^2(\mathbf{r},t)}{\omega_{p0}}\cdot\overrightarrow{\nabla}_{\mathbf{k}}\right]\frac{1}{2}\left\{(\tau_3+i\tau_2),W^{(0)}\right\}=0
\label{eqn:19}
\end{equation}
\begin{equation}
\left[\left(c^2\mathbf{k}^2+\tilde{\omega}_p^2(\mathbf{r},t)\right)\frac{(\tau_3+i\tau_2)}{2\omega_{p0}}+\omega_{p0}\tau_3,W^{(0)}\right]=0
\label{eqn:20}
\end{equation}
\end{subequations}

Our first goal now is to find a matrix operator that transforms the equation (\ref{eqn:20}), into a necessary condition for $W^{\prime}=U^{-1}W^{(0)}U$ to be diagonal. This will significantly reduce the amount of calculations needed to identify the form of the classical wave action, since the system will be reduced to two degrees of freedom. The matrix operator that performs this task is:
\begin{equation}
U(\zeta)=e^{-\tau_1\zeta}
\label{eqn:u}
\end{equation}
 where $\zeta=\frac{1}{2}\log\left[\frac{\omega(\mathbf{k},\mathbf{r},t)}{\omega_{p0}}\right]$, 
 and with $\omega(\mathbf{k},\mathbf{r},t)=\sqrt{\mathbf{k}^2c^2+\omega_{p0}^2+\tilde{\omega}_p^2(\mathbf{r},t)}$. 
On performing the similarity transformation induced by $U$ in Eq.~(\ref{eqn:20}), we obtain:
\begin{equation}
\omega(\mathbf{k},\mathbf{r},t)\left[\tau_3,W^{\prime}\right]=0
\label{eqn:21}
\end{equation}
where 
\begin{equation}
U^{-1}(\tau_3+i\tau_2)U=\frac{\omega_{p0}}{\omega(\mathbf{k},\mathbf{r},t)} (\tau_3+i\tau_2)
\end{equation}

From equation (\ref{eqn:21}) we can see that $W^{\prime}$ must be a linear combination of $\tau_0$ and $\tau_3$, since all the other Pauli matrices do  not commute with $\tau_3$. One can choose this linear combination to be parameterized by two real functions $f$ and $g$, such that:
\begin{equation}
W^{\prime}=\frac{f-g}{2}\tau_0+\frac{f+g}{2}\tau_3
\label{eqn:22}
\end{equation}
Performing the transformation (\ref{eqn:u}) in Eq.~(\ref{eqn:19}), with $W^{\prime}$ given by Eq.~(\ref{eqn:22}), 
the diagonal part leads to
\begin{subequations}
\begin{equation}
\partial_t f(\mathbf{k},\mathbf{r},t)+\frac{c^2\mathbf{k}}{\omega(\mathbf{k},\mathbf{r},t)}\cdot\overrightarrow{\nabla}_{\mathbf{r}}f(\mathbf{k},\mathbf{r},t)-\frac{1}{2}\frac{\overrightarrow{\nabla}_{\mathbf{r}}\tilde{\omega}_p^2(\mathbf{r},t)}{\omega_{p0}}\cdot\overrightarrow{\nabla}_{\mathbf{k}}f(\mathbf{k},\mathbf{r},t)=0
\label{eqn:23}
\end{equation}
\begin{equation}
\partial_t g(\mathbf{k},\mathbf{r},t)-\frac{c^2\mathbf{k}}{\omega(\mathbf{k},\mathbf{r},t)}\cdot\overrightarrow{\nabla}_{\mathbf{r}}g(\mathbf{k},\mathbf{r},t)+\frac{1}{2}\frac{\overrightarrow{\nabla}_{\mathbf{r}}\tilde{\omega}_p^2(\mathbf{r},t)}{\omega_{p0}}\cdot\overrightarrow{\nabla}_{\mathbf{k}}g(\mathbf{k},\mathbf{r},t)=0
\label{eqn:24}
\end{equation}
\end{subequations}

Equations (\ref{eqn:23}) and (\ref{eqn:24}) clearly resemble the classical wave action conservation for forward and backward radiation, respectively \cite{bib:mcdonald}. However to confirm this fact, the relation between $f$, $g$ and $W_{\Phi\Phi}^{(0)}$ needs to be established. On one hand, we know that $\mathrm{Re}(W)$ is $1/2(\tau_0W_0-i\tau_2W_2+\tau_3W_3)$. On the other hand, in the short wavelength limit, $\mathrm{Re}(W)$ is $W^{(0)}$. Thus, we conclude that:
\begin{equation}
W^{(0)}=\frac{1}{2}\left(\tau_0W_0^{(0)}-i\tau_2W_2^{(0)}+\tau_3W_3^{(0)}\right)
\label{eqn:temp3}
\end{equation}
Performing the transformation induced by $U$ in Eq.~(\ref{eqn:temp3}) we obtain:
\begin{equation}
2W^{\prime}=(f-g)\tau_0+(f+g)\tau_3=U^{-1}(\tau_0W_0^{(0)}-i\tau_2W_2^{(0)}+\tau_3W_3^{(0)})U
\label{eqn:25}
\end{equation}
which after some manipulation can be reduced to the set of algebraic equations:
\begin{subequations}
\begin{equation}
f-g=W_0^{(0)}
\end{equation}
\begin{equation}
f+g=\frac{R^2+1}{2 R}W_3^{(0)}+\frac{R^2-1}{2 R}W_2^{(0)}
\end{equation}
\begin{equation}
0=\frac{R^2+1}{2 R}W_2^{(0)}+\frac{R^2-1}{2 R}W_3^{(0)}
\end{equation}
\label{eqn:26}
\end{subequations}
where we have defined $R=\omega(\mathbf{k},\mathbf{r},t)/\omega_{p0}$. From Eqs. (\ref{eqn:26}), one infers:
\begin{equation}
f+g=\frac{\omega(\mathbf{k},\mathbf{r},t)}{\omega_{p0}}W_{\Phi\Phi}^{(0)}=\frac{c^2}{2\omega_{p0}}J(\mathbf{k},\mathbf{r},t)
\end{equation}
where $J(\mathbf{k},\mathbf{r},t)$ represents the classical wave action, as defined in ref. \cite{bib:mcdonald}. Thus, in the short wavelength limit, Eqs.~(\ref{eqn:23},\ref{eqn:24}) state that the classical wave action, $J(\mathbf{k},\mathbf{r},t)$, is conserved along the geometrical optics ray, independently of the adiabaticity condition \cite{bib:bellotti}.

To conclude the discussion of the short wavelength limit, we observe that when $\varepsilon\to0$ in Eq. (\ref{eqn:3}) $\chi$ is negligible, since reflected waves are exponentially small \cite{bib:mahony}. Dropping the small component $\chi$ in Eq.~(\ref{eqn:3}) and eliminating the $\omega_{p0}$ term by setting $\phi(\mathbf{r},t)=\tilde{\phi}(\mathbf{r},t)\exp{(-i\omega_{p0}t/\varepsilon)}$, yields the paraxial wave equation \cite{bib:lasers}:
\begin{equation}
2 i \omega_{p0}\varepsilon\partial_t \tilde{\phi}=-\varepsilon^2c^2\overrightarrow{\nabla}_{\mathbf{r}}^2\tilde{\phi}+\tilde{\omega}_p^2\tilde{\phi}
\end{equation}
This equation is formally equivalent to the Schr\"odinger equation, and the Wigner-Moyal formalism can be followed in a straightforward way. This is the approach followed by Besieris and Tappert \cite{bib:besieris1,bib:besieris}. As expected, this path also leads to classical wave action conservation in the same limit we have explored here, but fails to capture the dynamics of the backscattered component of the radiation field.

\section{\label{sec:five}Conclusions}
In order to establish a description of the propagation of electromagnetic waves in nonlinear dispersive media, including the two modes described by the wave equation, we have employed the Feshbach-Villars description of the Klein-Gordon field with a variable mass term, and the corresponding $2\times2$ Wigner matrix. The equations of motion for the Wigner matrix lead to a set of four coupled equations for four real phase-space densities, whose physical meanings are clearly identified. The general Boltzmann-Vlasov equations of motion account for the backscattered waves, and for the interference between forward and backward radiation. These equations replace the standard Wigner-Moyal description of classical fields \cite{bib:besieris}, generalizing it to the two mode problem with variable mass terms, and are formally equivalent to the full wave equation.

The short wavelength limit of the Boltzmann-Vlasov equations of motion was considered, and it was shown they reduce to uncoupled equations for the forward and backward radiation phase-space densities. In this limit, the standard results regarding wave action conservation, \cite{bib:mcdonald} are recovered.

The formalism introduced here provides a complete description for the electromagnetic field in nonlinear dispersive media, and opens the way to the description of broadband radiation driven parametric instabilities in nonlinear media. Applications of this formalism will be presented elsewhere.


\end{document}